\documentclass[aps,prl,twocolumn,superscriptaddress,amsfont,graphicx,nofootinbib,preprintnumbers]{revtex4}

\usepackage{color,graphicx,epsfig}
\usepackage{ifpdf}
\usepackage{amsmath}
\usepackage{bm}
\usepackage{color}
\usepackage[english]{babel}
\usepackage{graphicx}
\usepackage{amsfonts}
\usepackage{amssymb}
\usepackage{braket}
\usepackage{hyperref}
\usepackage[utf8]{inputenc}

\definecolor{nicered}{rgb}{0.7,0.1,0.1}
\definecolor{nicegreen}{rgb}{0.1,0.5,0.1}
\hypersetup{colorlinks,citecolor= nicegreen,linkcolor= nicered}

\newcommand{\beq}{\begin{equation}}
\newcommand{\eeq}{\end{equation}}
\newcommand{\bea}{\begin{eqnarray}}
\newcommand{\eea}{\end{eqnarray}}
\newcommand{\kc}{\kappa_c}
\newcommand{\kb}{\kappa_b}

\newcommand{\kq}{\kappa_Q}

\newcommand{\mq}{m_Q}
\newcommand{\mqsq}{m_Q^2}
\newcommand{\pperp}{p_\perp}
\newcommand{\pperpsq}{p_\perp^2}
\newcommand{\yqsm}{y_Q^{\rm SM}}
\newcommand{\gluon}{gg \to hj}
\newcommand{\quark}{gQ \to hQ,\, Q\bar{Q}\to hg}
\def \bm#1{\mbox{\boldmath$#1$\unboldmath}} 

\begin{document}

\def\Oxford{Rudolf Peierls Centre for Theoretical Physics, University of Oxford OX1 3NP Oxford, United Kingdom}
\def\CERN{CERN, Theoretical Physics Department, CH-1211 Geneva 23, Switzerland}
\def\LAPTh{LAPTh, Universit\'e Savoie Mont Blanc, CNRS, B.P.110, Annecy-le-Vieux F-74941, France}

\preprint{OUTP-16-18P}
\preprint{CERN-TH-2016-136}
\preprint{LAPTH-026/16}

\title{Constraining Light-Quark Yukawa Couplings from Higgs Distributions}

\author{Fady Bishara}
\email[Electronic address:]{fady.bishara@physics.ox.ac.uk} 
\affiliation{\Oxford}
\author{Ulrich Haisch}
\email[Electronic address:]{Ulrich.Haisch@physics.ox.ac.uk} 
\affiliation{\Oxford}
\affiliation{\CERN}
\author{Pier Francesco Monni} 
\email[Electronic address:]{Pier.Monni@physics.ox.ac.uk} 
\affiliation{\Oxford}
\author{Emanuele Re}
\email[Electronic address:]{emanuele.re@lapth.cnrs.fr} 
\affiliation{\LAPTh}

\begin{abstract} 
We propose a novel strategy to constrain the bottom and charm Yukawa couplings by exploiting LHC measurements of transverse momentum distributions in Higgs production. Our method does not rely on the reconstruction of exclusive final states or heavy-flavour tagging. Compared to other proposals, it leads to an enhanced sensitivity to the Yukawa couplings due to distortions of the differential Higgs spectra from emissions which either probe quark loops or are associated with quark-initiated production.  We derive constraints using data from LHC Run I, and we explore the prospects of our method at future~LHC runs.  Finally, we comment on the possibility of bounding the strange Yukawa coupling.
\end{abstract}


\maketitle

{\bf Introduction.} The discovery of a spin-0 resonance and the measurement of its couplings to the standard model (SM) gauge bosons~\cite{Aad:2015gba,Khachatryan:2014jba} have established that the dominant source of electroweak symmetry breaking is the vacuum expectation value (VEV) of a scalar field. In the SM this Higgs VEV is also responsible for giving mass to all charged fermions and the LHC Run~I measurements support this simple picture in the case of the top and bottom Yukawa couplings $y_t$ and $y_b$. Direct measurements of the charm Yukawa coupling are on the other hand not available at present, and it has been common lore (see~e.g.~\cite{Berger:2002vs,Peskin:2012we}) that extractions of $y_c$ can only be performed with a few-percent uncertainty at an~$e^+ e^-$ machine such as the ILC~\cite{Ono:2013sea}.

Only recently it has been realised that gaining direct access to $y_c$ without the ILC is possible as in its high-luminosity run the LHC (HL-LHC) will serve as a Higgs factory producing around $1.7 \cdot 10^8$ Higgs bosons per experiment with~$3 \, {\rm ab}^{-1}$ of integrated luminosity~\cite{Dawson:2013bba}. In fact, several different strategies have been proposed to constrain modifications $\kc = y_c/y_c^{\rm SM}$.\footnote{Here $\yqsm = \sqrt{2} \hspace{0.25mm} \mq/v$ with $v \simeq 246 \, {\rm GeV}$ and $\mq$ is a $\overline {\rm MS}$ mass renormalised at the scale $m_h/2$. In our numerical analysis, we employ $y_b^{\rm SM} = 1.9 \cdot 10^{-2}$, $y_c^{\rm SM} = 4.0 \cdot 10^{-3}$ and $y_s^{\rm SM} = 3.3 \cdot 10^{-4}$.}  A first way to probe~$\kc$ consists in searching for the exclusive decay~$h \to J/\psi \gamma$~\cite{Bodwin:2013gca,Kagan:2014ila,Koenig:2015pha}. While reconstructing the $J/\psi$ via its di-muon decay leads to a clean experimental signature, the small branching ratio, ${\rm Br} \left ( h \to J/\psi \gamma \to \mu^+ \mu^- \gamma \right ) = 1.8 \cdot 10^{-7}$, implies that only 30 signal events can be expected at each experiment. This makes a detection challenging given the large continuous background due to QCD production of charmonia and a jet faking a photon \cite{Aad:2015sda,Perez:2015lra}. Search strategies with larger signal cross sections are $p p \to W/Z \hspace{0.25mm} h \, (h \to c \bar c)$~\cite{Delaunay:2013pja,Perez:2015lra,Perez:2015aoa} and $p p \to h c$~\cite{Brivio:2015fxa}. These strategies, however, rely on charm tagging~($c$-tagging) algorithms~\cite{ATL-PHYS-PUB-2015-001,SethMoortgat} which are currently inefficient.  Given these limitations, it is important to devise another independent procedure that neither suffers from a small signal rate nor depends on the $c$-tagging performance. In this letter, we will present a method that relies on the measurements of transverse momentum distributions of Higgs plus jets events. This signature receives contributions from gluon fusion~($\gluon$) and quark-initiated production~($\quark$). In the $\gluon$ channel, the Higgs is produced through quark loops that are probed by real emissions in specific kinematic regimes. In particular, when emissions have a transverse momentum~$\pperp$ in the range $\mq \ll \pperp \ll m_h$, with~$\mq$ being the internal quark mass, the leading-order (LO) cross section features double logarithms of the form~\cite{Baur:1989cm}
\begin{equation}
\label{eq:largelogarithms} 
\kq \; \frac{\mqsq}{m_h^2} \, \ln^2 \left ( \frac{\pperpsq}{\mqsq} \right) \,, 
\end{equation}
due to the interference between the $Q$-mediated and the top-mediated contributions. These logarithms dynamically enhance the dependence on the Yukawa modification $\kq$. The differential cross section of $gg\to h$ receives radiative corrections which contain up to two powers of the logarithm $\ln \left (\pperpsq/\mqsq \right )$ for each extra power of the strong coupling constant~$\alpha_s$. If instead the Higgs is produced in $\quark$, the resulting LO differential cross section scales as $\kq^2$ (this channel therefore dominates in the large-$\kq$ regime that is relevant for first generation quarks~\cite{Soreq:2016rae}), with an additional suppression factor of ${\cal O} (\alpha_s/\pi)$ for each initial-state sea-quark parton distribution function~(PDF) which is generated perturbatively via gluon splitting. Owing to the different Lorentz structure of the amplitudes in the $m_Q\to 0$ limit, the $\gluon$ and $\quark$ processes do not interfere at ${\cal O} (\alpha_s^2)$. This ensures that no terms scaling linearly in~$\kq$ are present in the $\quark$ channels at this order.

The sensitivity to $y_Q$ that arises from the interplay between the different production modes can be studied by means of the differential spectra of the Higgs boson and jets transverse momentum (henceforth generically denoted by $p_T$) in the moderate-$p_T$ region. In fact, the double logarithms can be numerically large for transverse momenta $p_T \lesssim m_h/2$. This partly compensates for the quadratic mass suppression~$\mqsq/m_h^2$ appearing in~(\ref{eq:largelogarithms}). As a result of the logarithmic sensitivity and of the $\kq^2$ dependence in quark-initiated production, one expects deviations of several percent in the~$p_T$ spectra in Higgs production for ${\cal O} (1)$ modifications of $\kq$. In the SM, the light-quark effects are small. Specifically, in comparison to the Higgs effective field theory~(HEFT) prediction, in $\gluon$ the bottom contribution has an effect of around~$-5\%$ on the differential distributions while the impact of the charm quark is at the level of $-1\%$. Likewise, the combined $\quark$ channels~(with $Q=b,c$) lead to a shift of roughly~$2\%$. Precision measurements of the Higgs distributions for moderate $p_T$ values combined with precision calculations of these observables are thus needed to probe~${\cal O} (1)$ deviations in $y_b$ and~$y_c$. Achieving such an accuracy is both a theoretical and experimental challenge, but it seems possible in view of foreseen advances in higher-order calculations and the large statistics expected at future LHC~runs.

{\bf Theoretical framework.}  Our goal is to explore the sensitivity of the Higgs-boson~($p_{T,h}$) and leading-jet~($p_{T,j}$) transverse momentum distributions in inclusive Higgs production to simultaneous modifications of the light Yukawa couplings. We consider final states where the Higgs boson decays into a pair of gauge bosons. To avoid sensitivity to the modification of the branching ratios, we normalise the distributions to the inclusive cross section. The effect on branching ratios can be included in the context of a global analysis, jointly with the method proposed here.

The $\gluon$ channel was analysed in depth in the HEFT framework where one integrates out the dominant top-quark loops and neglects the contributions from lighter quarks. While in this approximation the two spectra and the total cross section were studied extensively, the effect of lighter quarks is not yet known with the same precision for $p_T \lesssim m_h/2$. Within the SM, the LO distribution for this process was derived long ago~\cite{Ellis:1987xu,Baur:1989cm}, and the next-to-leading-order~(NLO) corrections to the total cross section were calculated in~\cite{Spira:1995rr,Spira:1997dg,Harlander:2005rq,Anastasiou:2006hc,Aglietti:2006tp}. In the context of analytic resummations of the Sudakov logarithms~$\ln \left(p_T/m_h \right)$, the inclusion of mass corrections to the HEFT were studied both for the $p_{T,h}$ and~$p_{T,j}$ distributions~\cite{Mantler:2012bj,Grazzini:2013mca,Banfi:2013eda}. More recently, the first resummations of some of the leading logarithms~\eqref{eq:largelogarithms} were accomplished both in the abelian~\cite{Melnikov:2016emg} and in the high-energy~\cite{Caola:2016upw} limit. The reactions $\quark$ were computed at NLO~\cite{Campbell:2002zm,Harlander:2010cz} in the five-flavour scheme that we employ here, and the resummation of the logarithms~$\ln \left(p_{T,h}/m_h \right)$ in $Q\bar{Q}\to h$ was also performed up to next-to-next-to-leading-logarithmic (NNLL) order~\cite{Harlander:2014hya}.

In the case of $\gluon$, we generate the LO spectra with \texttt{MG5$_{ }$aMC@NLO}~\cite{Alwall:2014hca}. We also include NLO corrections to the spectrum in the HEFT~\cite{deFlorian:1999zd,Ravindran:2002dc,Glosser:2002gm} using~\texttt{MCFM}~\cite{Campbell:2015qma}. The total cross sections for inclusive Higgs production are obtained from \texttt{HIGLU}~\cite{Spira:1995mt}, taking into account the
NNLO corrections in the~HEFT~\cite{Harlander:2002wh,Anastasiou:2002yz,Ravindran:2003um}. Sudakov logarithms $\ln \left (p_T/m_h \right)$ are resummed up to NNLL order both for $p_{T,h}$~\cite{Bozzi:2003jy,Becher:2010tm,Monni:2016ktx} and~$p_{T,j}$~\cite{Banfi:2012jm,Becher:2013xia,Stewart:2013faa}, treating mass corrections following~\cite{Banfi:2013eda}. The latter effects will be significant, once the spectra have been precisely measured down to $p_T$ values of~${\cal O} (5 \, {\rm GeV})$. The $\quark$ contributions to the distributions are calculated at NLO with \texttt{MG5$_{ }$aMC@NLO}~\cite{Wiesemann:2014ioa} and cross-checked against~\texttt{MCFM}. The obtained events are showered with~\texttt{PYTHIA~8.2}~\cite{Sjostrand:2014zea} and jets are reconstructed with the anti-$k_t$ algorithm~\cite{Cacciari:2008gp} as implemented in \texttt{FastJet}~\cite{Cacciari:2011ma} using $R = 0.4$ as a radius parameter.

\begin{figure}[!t]
\begin{center}
\includegraphics[width=0.4\textwidth]{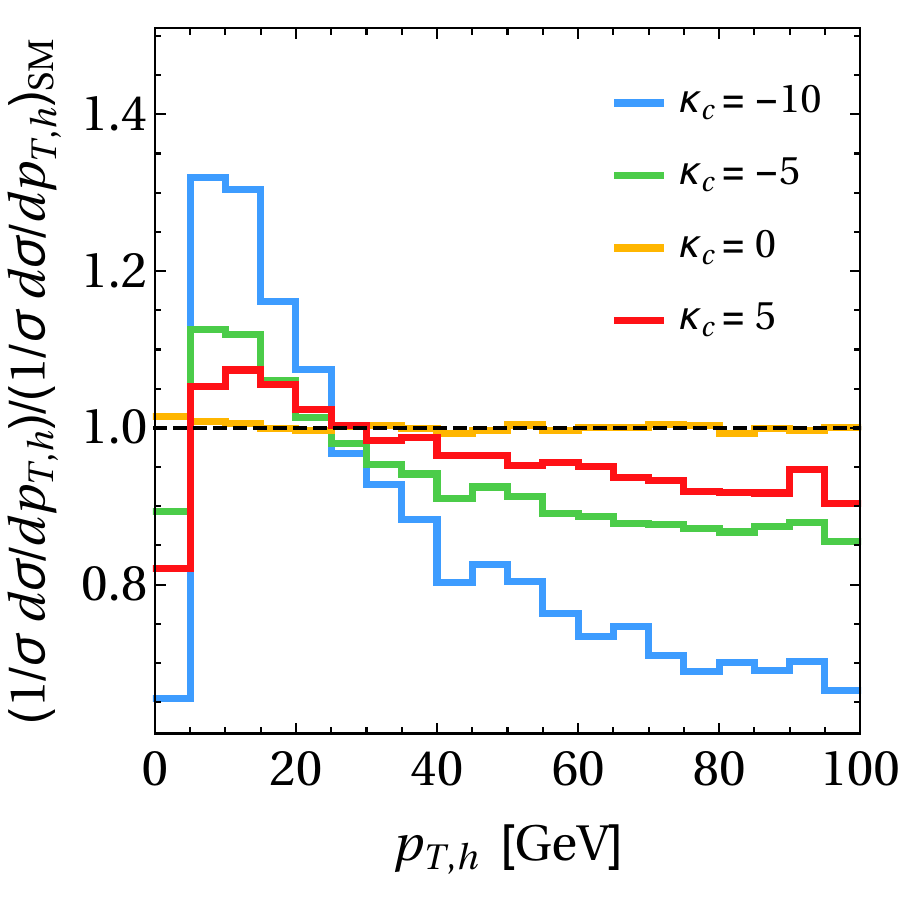} \vspace{0mm} 
\caption{\label{fig:spectra} The  normalised $p_{T,h}$ spectrum of inclusive Higgs production at $\sqrt{s} = 8 \, {\rm TeV}$ divided by the SM prediction for different values of~$\kc$. Only $\kc$ is modified, while the remaining Yukawa couplings are kept at their SM values.}
\end{center}
\end{figure}

Our default choice for the renormalisation ($\mu_R$), factorisation ($\mu_F$) and the resummation ($Q_R$, for $\gluon$) scales is $m_h/2$.  Perturbative uncertainties are estimated by varying $\mu_R$, $\mu_F$ by a factor of two in either direction while keeping $1/2 \leq \mu_R/\mu_F \leq 2$. In addition, for the $\gluon$ channel, we vary $Q_R$ by a factor of two while keeping $\mu_R=\mu_F=m_h/2$. The final total theoretical errors are then obtained by combining the scale uncertainties in quadrature with a $\pm 2\%$ relative error associated with PDFs and~$\alpha_s$ for the normalised distributions. We stress that the normalised distributions used in this study are less sensitive to PDFs and~$\alpha_s$ variations, therefore the above $\pm 2\%$ relative uncertainty is a realistic estimate. We obtain the relative uncertainty in the SM and then assume that it does not depend on $\kq$. While this is correct for the $\quark$ channels, for the $\gluon$ production a good assessment of the theory uncertainties in the large-$\kq$ regime requires the resummation of the logarithms in~\eqref{eq:largelogarithms}. First steps in this direction were taken in~\cite{Melnikov:2016emg, Caola:2016upw}. 

On the other hand, in the small-$\kq$ regime that will be probed at future runs of the LHC, the distribution is dominated by the $\gluon$ channel. For small values of $\kq$ the $\ln \left (p_T^2/\mqsq \right )$ terms are of moderate size and a good assessment of these effects comes from the NLO calculation of mass corrections in $\gluon$~\cite{Melnikov:2016qoc,Melnikov:2017pgf,Lindert:2017pky}. Furthermore, achieving a perturbative uncertainty of a few percent in the considered $p_T$ region would also require improving the accuracy of the resummed $\ln \left(p_{T}/m_h \right)$ terms beyond NNLL. Progress in this direction~\cite{Becher:2013xia,Li:2016ctv} suggests that this will be achieved in the near future. Incorporating higher-order corrections to the full SM process will both reduce the theoretical uncertainties and improve the sensitivity to $\kq$.

Figure~\ref{fig:spectra} illustrates the impact of the Yukawa modification $\kc$ on the normalised~$p_{T,h}$ spectrum in inclusive Higgs production. The results are divided by the SM prediction and correspond to~$pp$ collisions at a centre-of-mass energy ($\sqrt{s}$) of~$8 \, {\rm TeV}$,\footnote{The ratio of the $p_{T,h}$ spectra to the SM prediction at $\sqrt{s} = 13 \, {\rm TeV}$ is slightly harder than the $\sqrt{s} = 8 \, {\rm TeV}$ counterpart, which enhances the sensitivity to $\kb$ and $\kc$ at ongoing and upcoming LHC runs as well as possible future hadron colliders at higher energies.} central choice of scales and {\tt MSTW2008NNLO} PDFs~\cite{Martin:2009iq}. Notice that for $p_{T,h} \gtrsim 50 \, {\rm GeV}$, the asymptotic behaviour~\eqref{eq:largelogarithms} breaks down and
consequently the $\quark$ channels control the shape of the~$p_{T,h}$ distributions.

We stress that for the $p_{T,h}$ distribution, non-perturbative corrections are small and in the long run, $p_{T,h}$ will be measured to lower values than $p_{T,j}$. While the latter currently gives comparable sensitivity, it is mandatory to study $p_{T,h}$ to maximise the constraints on $\kq$ in future LHC runs. Therefore, we use $p_{T,h}$ in the rest of this letter.

\begin{figure}[!t]
\begin{center}
\includegraphics[width=0.9\columnwidth]{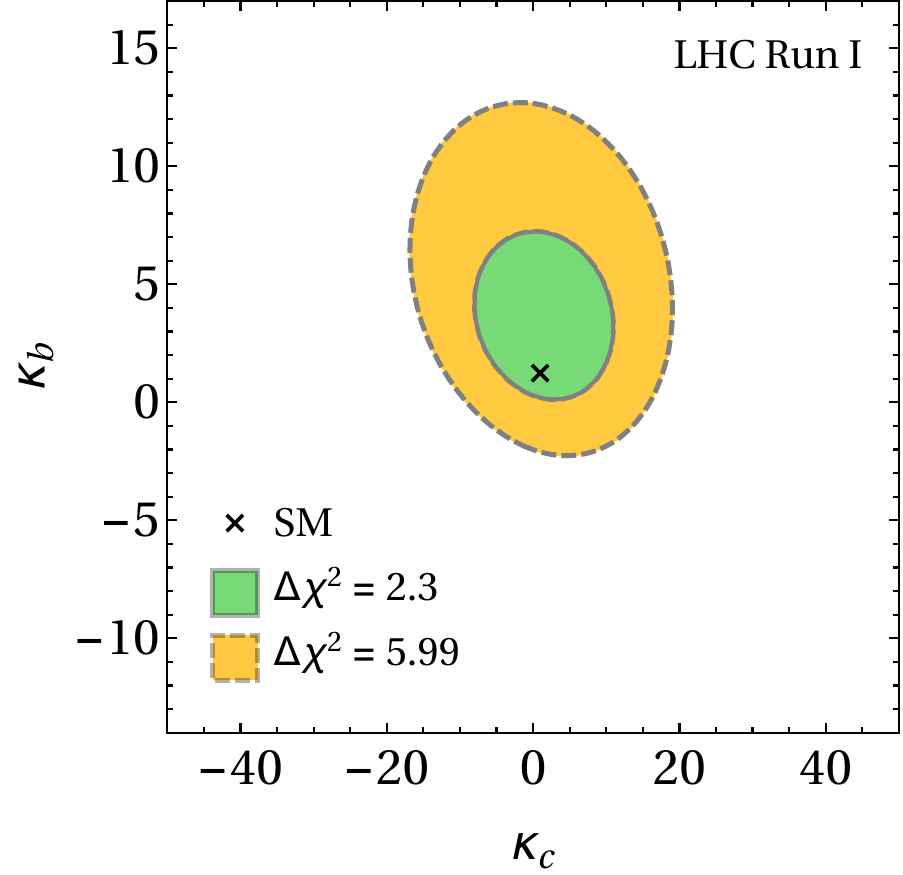}  
\vspace{-2mm}
\caption{\label{fig:present} The $\Delta\chi^2=2.3$  and $\Delta\chi^2=5.99$ regions in the $\kc \hspace{0.25mm}$--$\hspace{0.25mm} \kb$ plane following from the combination of the  ATLAS measurements of the normalised $p_{T,h}$ distribution in  the $h \to \gamma \gamma$ and $h \to Z Z^\ast \to 4 \ell$ channels. The SM point is indicated by the black cross.}
\end{center}
\end{figure}

{\bf Current constraints.} At $\sqrt{s} = 8 \, {\rm TeV}$, the ATLAS and CMS collaborations have measured the $p_{T,h}$ and $p_{T,j}$ spectra in the $h \to \gamma \gamma$~\cite{Aad:2014lwa, Khachatryan:2015rxa}, $h \to Z Z^\ast \to 4 \ell$~\cite{Aad:2014tca, Khachatryan:2015yvw} and $h \to W W^\ast \to e\mu \nu_e \nu_\mu$~\cite{Aad:2016lvc,Khachatryan:2016vnn} channels, using around $20 \, {\rm fb}^{-1}$ of data in each case. To derive constraints on~$\kb$ and~$\kc$, we harness the normalised~$p_{T,h}$ distribution in inclusive Higgs production~\cite{Aad:2015lha}. This spectrum is obtained by ATLAS from a combination of $h \to \gamma \gamma$ and $h \to Z Z^\ast \to 4 \ell$ decays, and represents at present the most precise measurement of the differential inclusive Higgs cross section. In our $\chi^2$ analysis, we include the first seven bins in the range $p_{T,h} \in [0,100]$\,GeV whose experimental uncertainty is dominated by the statistical error. This data is then compared to the theoretical predictions for the inclusive~$p_{T,h}$ spectrum described in the previous section. We assume that all the errors are Gaussian in our fit. The bin-to-bin correlations in the theoretical normalised distributions are obtained by assuming that the bins of the unnormalised distributions are uncorrelated and modelled by means of linear error propagation. This accounts for the dominant correlations in normalised spectra. For the data, we used the correlation matrix of \cite{Aad:2015lha}.

Figure~\ref{fig:present} displays the $\Delta\chi^2=2.3$  and $\Delta\chi^2=5.99$ contours (corresponding to a 68\% and 95\% confidence level~(CL) for a Gaussian distribution) in the $\kc \hspace{0.25mm}$--$\hspace{0.25mm} \kb$ plane. We profile over $\kb$ by means of the profile likelihood ratio~\cite{Cowan:2010js} and obtain the following 95\%~CL bound
\beq \label{eq:kappacboundpresent} 
\kc \in [-16,18] \quad (\text{LHC Run I}) \,.  
\eeq 
Our limit is significantly stronger than the bounds from exclusive $h \to J/\psi \gamma$ decays~\cite{Aad:2015sda}, a recast of $h \to b \bar b$ searches and the measurements of the total Higgs width~\cite{Khachatryan:2014jba, Aad:2014aba}, which read $|\kc| \lesssim 429$~\cite{Koenig:2015pha}, $|\kc| \lesssim 234$ and $|\kc| \lesssim 130$~\cite{Perez:2015aoa}, respectively. It is however not competitive with the bound~$|\kc| \lesssim 6.2$ from a global analysis of Higgs data~\cite{Perez:2015aoa}, which introduces additional model dependence.

Turning our attention to the allowed modifications of the bottom Yukawa coupling, one observes that our proposal leads to $\kb \in [-3,15]$. This limit is thus significantly weaker than the constraints from the LHC Run~I measurements of $p p \to W/Z \hspace{0.25mm} h \, (h \to b \bar b)$, $p p \to t \bar t \hspace{0.25mm} h \, (h \to b \bar b)$ and $h \to b \bar b$ in vector boson fusion that already restrict the relative shifts in~$y_b$ to around $\pm50\%$~\cite{Aad:2015gba,Khachatryan:2014jba}.

{\bf Future prospects.} As a result of the expected reduction of the statistical uncertainties for the $p_{T, h}$ spectrum at the LHC, the proposed method will be limited by systematic uncertainties in the long run. Recent studies by CMS~\cite{CMS-DP-2016-064} show that the residual experimental systematic uncertainty will be reduced to the level of a few percent at the HL-LHC. Therefore, it is natural to study the prospects of the method in future scenarios assuming a reduced theory uncertainty given that this error may become the limiting factor.

In order to investigate the future prospects of our method, we need a more precise assessment of the non-perturbative corrections to the $p_{T,h}$ distribution. To estimate these effects, we used \texttt{MG5aMC@NLO} and \texttt{POWHEG}~\cite{Alioli:2008tz} showered with \texttt{Pythia~8.2} and found that the corrections can reach up to $2\%$ in the relevant $p_{T,h}$ region. This finding agrees with recent analytic studies of non-perturbative corrections to $p_{T,h}$ (see e.g.~\cite{Becher:2012yn}). With improved perturbative calculations, a few-percent accuracy in this observable will therefore be reachable. 

We study two benchmark cases. Our LHC Run II scenario employs~$0.3 \, {\rm
ab}^{-1}$ of integrated luminosity and assumes a systematic error of~$\pm 3\%$
on the experimental side and a total theoretical uncertainty of $\pm 5\%$. This
means that we envision that the non-statistical uncertainties present at LHC
Run~I can be halved in the coming years, which seems plausible. Our HL-LHC
scenario instead uses~$3 \, {\rm ab}^{-1}$ of data and foresees a reduction of
both systematic and theoretical errors by another factor of two, leading to
uncertainties of~$\pm 1.5\%$ and~$\pm 2.5\%$, respectively. The last scenario
is illustrative of the reach that can be achieved with improved theory
uncertainties. Alternative theory scenarios are discussed in the appendix. In
both benchmarks, we employ $\sqrt{s} = 13 \, {\rm TeV}$ and the {\tt
PDF4LHC15\_nnlo\_mc}
set~\cite{Butterworth:2015oua,Dulat:2015mca,Harland-Lang:2014zoa,Ball:2014uwa},
consider the range $p_T \in [0, 100] \, {\rm GeV}$ in bins of $5 \, {\rm GeV}$,
and take into account $h \to \gamma \gamma$, $h \to Z Z^\ast \to 4 \ell$ and $h \to W W^\ast \to 2 \ell 2 \nu_\ell$. We assume that future measurements will be centred around the~SM predictions. These channels sum to a branching ratio of $1.2\%$, but given the large amount of data the statistical errors per bin will be at the $\pm 2\%$~($\pm 1\%$) level in our LHC Run~II (HL-LHC) scenario. We model the correlation matrix as in the $8$\,TeV case.

\begin{figure}[!t]\centering
\includegraphics[width=0.9\columnwidth]{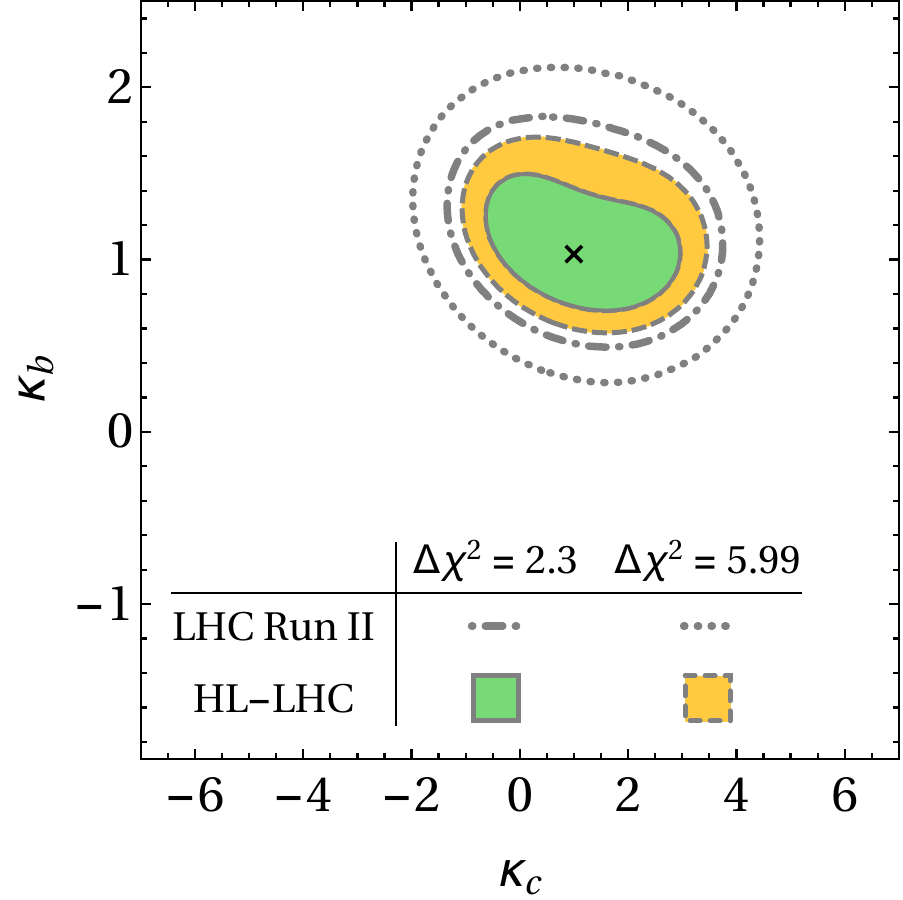} \vspace{-2mm}
\caption{Projected future constraints in the $\kc \hspace{0.25mm}$--$\hspace{0.25mm} \kb$ plane. The SM point is indicated by the black cross. The figure shows our projections for the LHC Run II (HL-LHC) with $0.3 \, {\rm ab}^{-1}$ ($3 \, {\rm ab}^{-1}$) of integrated luminosity at $\sqrt{s} = 13 \, {\rm TeV}$. The remaining assumptions entering our future predictions are detailed in the main text.}
\label{fig:future} 
\end{figure}

The results of our $\chi^2$ fits are presented in Figure~\ref{fig:future}, showing the constraints in the $\kc \hspace{0.25mm}$--$\hspace{0.25mm} \kb$ plane. The unshaded contours refer to the LHC Run II scenario with the dot-dashed (dotted) lines corresponding to $\Delta\chi^2=2.3\;(5.99)$. Analogously, the shaded contours with the solid (dashed) lines refer to the HL-LHC. By profiling over $\kb$, we find in the LHC~Run~II scenario the following 95\%~CL bound on the $y_c$ modifications 
\beq \label{eq:futuremarginalisation1}
\kc \in [-1.4,3.8]  \quad  (\text{LHC Run II})\,,
\eeq
while the corresponding HL-LHC bound reads 
\beq \label{eq:futuremarginalisation2}
\kc \in [-0.6,3.0]  \quad  (\text{HL-LHC})\,.
\eeq
These limits compare well not only with the projected reach of other proposed strategies but also have the nice feature that they are controlled by the size of systematic uncertainties that can be reached in the future.
Also, at future LHC runs our method will allow one to set relevant bounds on the modifications of $y_b$. For instance, in the HL-HLC scenario we obtain $\kb \in [0.7,1.6]$ at 95\%~CL.

Finally, we also explored the possibility of constraining modifications of the strange Yukawa coupling. Under the assumption that $y_b$ is SM-like but profiling over $\kc$, we find that at the HL-LHC one should have a sensitivity to~$y_s$ values of around 30 times the SM expectation. Measurements of exclusive $h \to \phi \gamma$ decays are expected to have a reach that is weaker than this by a factor of order 100~\cite{Perez:2015lra}.

{\bf Conclusions.} In this letter, we have demonstrated that the normalised~$p_T$ distribution of the Higgs or of jets recoiling against it, provide sensitive probes of the bottom, charm and strange Yukawa couplings.  Our new proposal takes advantage of the fact that the differential Higgs plus jets cross section receives contributions from the channels $\gluon$, $\quark$ that feature two different functional dependences on $\kq$. We have shown that in the kinematic region where the transverse momentum~$\pperp$ of emissions is larger than the relevant quark mass $\mq$, but smaller than the Higgs mass~$m_h$, both effects can be phenomenologically relevant and thus their interplay results in an enhanced sensitivity to  $\kq$. This feature allows one to obtain unique constraints on~$y_b$, $y_c$ and $y_s$ at future LHC runs.

We derived constraints in the $\kc \hspace{0.25mm}$--$\hspace{0.25mm} \kb$ plane that arise from LHC~Run~I data and provided sensitivity projections of our method in future runs of the LHC. Our results are obtained under the assumption that physics beyond the SM only alters the shape of the distributions via changes of $y_c$ and $y_b$, while modifications of the effective $ggh$ coupling that are induced by loops of new heavy states are not considered. Since effects of the latter type mainly affect the total Higgs production cross section, they largely cancel in the normalised $p_{T, h}$ spectrum for low and moderate values of the transverse momentum. The remaining model dependence, due to interference of heavy new physics with the light-quark loops, is subleading in the relevant regions of the $\kc \hspace{0.25mm}$--$\hspace{0.25mm} \kb$ plane.

Under reasonable assumptions about theoretical progress in the calculation of Higgs plus jets production and the precision of forthcoming experimental measurements we have shown that, at the HL-LHC, it is possible to obtain a limit of $\kc \in [-0.6,3.0]$ at 95\%~CL using our method alone. Modifications of this order can be realised in some models of flavour (see e.g.~\cite{Bishara:2015cha,Bauer:2015fxa}). The fact that our procedure is neither afflicted by a small signal rate nor depends on the performance of heavy-flavour tagging makes it highly complementary to extractions of~$y_{c}$ via $h \to J/\psi \gamma$, $pp \to W/Z \hspace{0.25mm} h \, (h \to c\bar c)$ and $pp \to hc$. In the case of the strange Yukawa coupling, we found that precision measurements of the Higgs $p_T$ distributions have a sensitivity to $y_s$ values of about 30 times the SM expectation, which exceeds the HL-LHC reach in $h \to \phi \gamma$ by about two orders of magnitude. 

\begin{acknowledgments} 
{\bf Acknowledgments.}  We express our gratitude to Gavin~Salam for drawing our attention to the quark-initiated contributions and pointing out that deriving bounds on the strange Yukawa coupling might be feasible. We thank Chris~Hays for helpful discussions concerning the ultimate experimental precision that measurements of $p_T$ spectra in Higgs production may reach at the HL-LHC, and Paolo~Torrielli for prompt help with~\texttt{MG5$_{ }$aMC@NLO}.  We are grateful to Florian~Bernlochner,  Chris~Hays, Gavin~Salam, Giulia~Zanderighi and Jure~Zupan for their valuable comments on the manuscript. UH acknowledges the hospitality and support of the CERN Theoretical Physics Department. 
\end{acknowledgments}

\section*{Appendix: A study of the dependence of the $\bm{\kappa_c}$ and $\bm{\kappa_b}$ bounds on the projected
systematic uncertainties}
\makeatletter

In this appendix we discuss in more detail the prospects of the proposed method
at the HL-LHC with $3 \, {\rm ab}^{-1}$ of integrated luminosity. In
particular, we examine how different assumptions about the experimental and
theoretical systematic uncertainties alter the resulting constraints on the
modifications $\kc$ and $\kb$ of the charm and bottom Yukawa couplings.

The HL-LHC projections  obtained in our letter assume an experimental systematic uncertainty of $1.5\%$ and a theoretical systematic error of $2.5\%$. This scenario illustrates the LHC reach based on an optimistic, but not unrealistic improvement  on both the experimental and theoretical side. Since it is difficult to forecast the precise figures for the  experimental and theoretical errors in the HL-LHC environment, it is interesting to explore the impact that variations of the systematic uncertainties have on the constraints on $\kc$ and $\kb$. 

\begin{table}[t!]
\centering
\begin{tabular}{c|c|c|c}
& \, experimental [\%] \, & \, theoretical [\%] \, & $\kappa_c \in$\\
\hline
\, $S_1$ \, & $1.5$ & $2.5$ & \, [-0.6, 3.0] \, \\
\, $S_2$ \, & $3.0$ & $2.5$ & \, [-0.9, 3.3] \, \\
\, $S_3$ \, & $1.5$ & $5.0$ & \, [-1.2, 3.6] \, \\
\, $S_4$ \, & $3.0$ & $5.0$ & \, [-1.3, 3.7] \, 
\end{tabular}
\caption{Experimental (second column) and theoretical~(third column) systematic uncertainties on the normalised Higgs transverse momentum ($p_{T, h}$) spectrum in our four uncertainty scenarios. The corresponding 95\%~CL constraints on $\kappa_c$ are also shown (fourth column) assuming $3 \, {\rm ab}^{-1}$ of data. }
\label{tab:errors}
\end{table}

In the following we study  four different uncertainty scenarios. They are described in Table~\ref{tab:errors}. Scenario~$S_1$ is the one employed in our letter to obtain the HL-LHC projections, while the systematical uncertainties of $S_4$ correspond to those used in the LHC~Run~II forecast. 

\begin{figure}[h]
\centering
 \includegraphics[width=0.9\columnwidth]{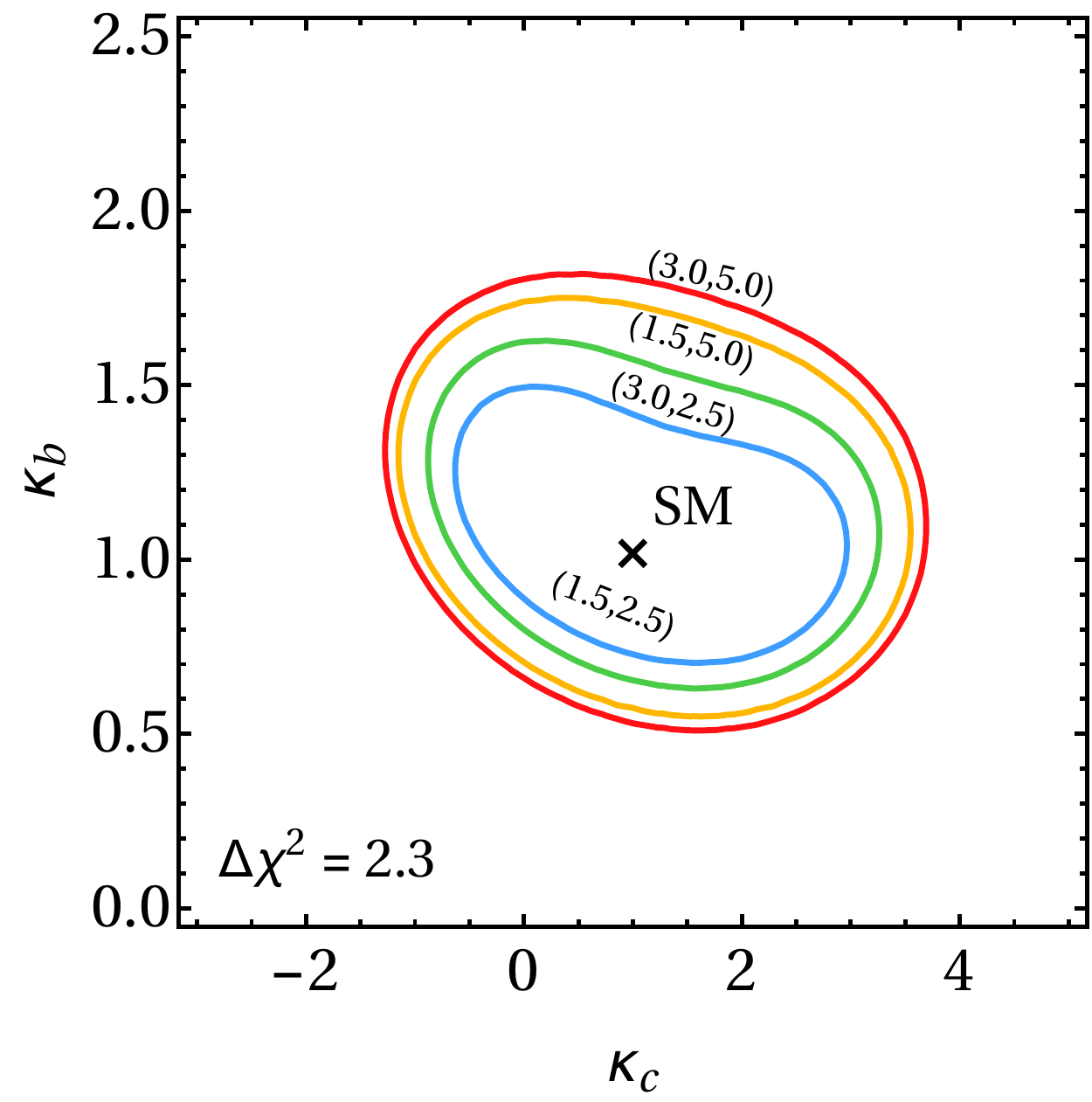}
 
 \vspace{4mm}
 
  \includegraphics[width=0.9\columnwidth]{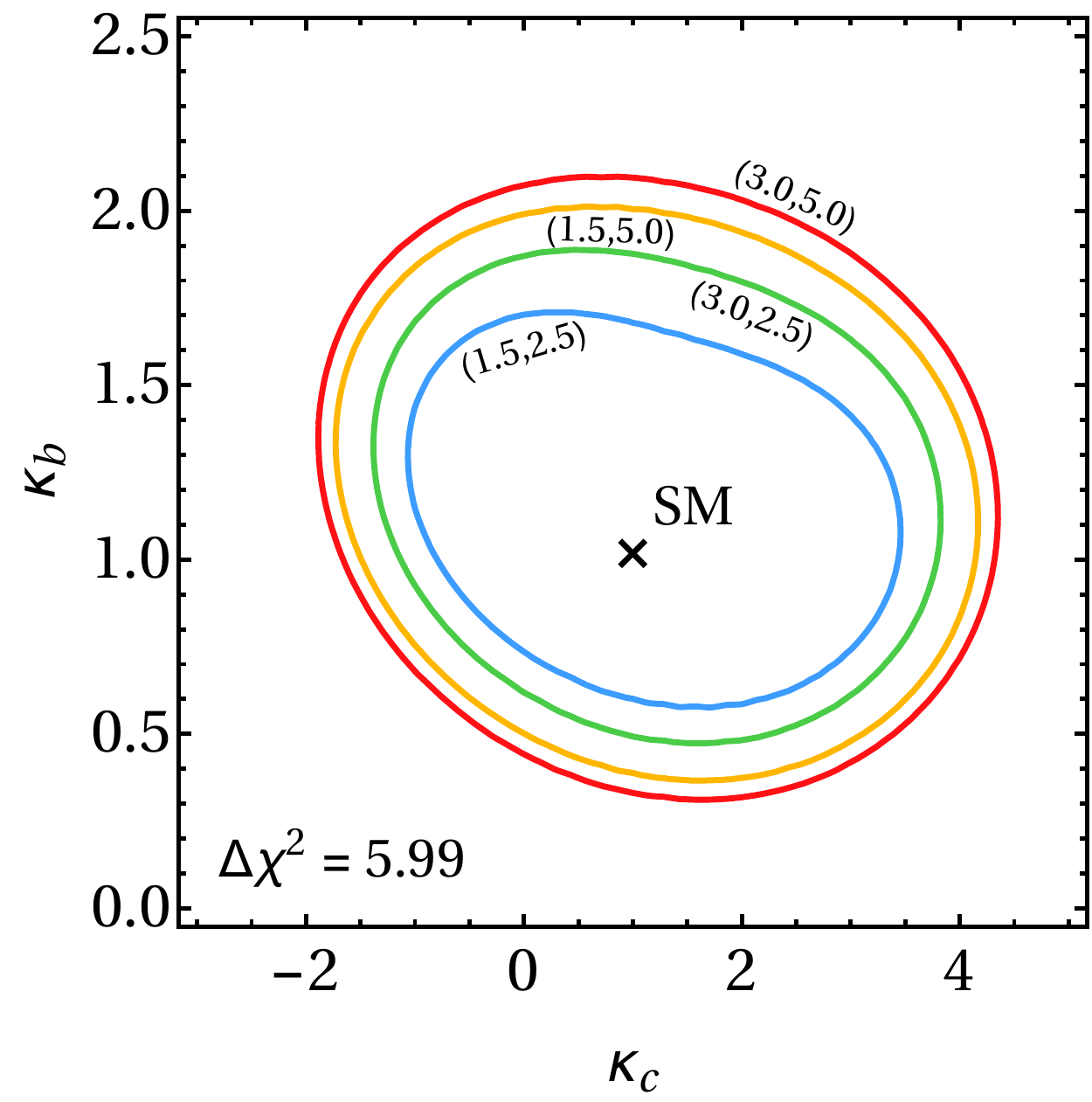}
  \caption{Projected 68\%~CL (upper plot) and 95\%~CL (lower plot)  constraints in the $\kc \hspace{0.25mm}$--$\hspace{0.25mm} \kb$ plane corresponding to the uncertainty scenarios of Table~\ref{tab:errors}. The numbers in brackets indicate the systematic experimental and theoretical uncertainty, respectively.}
 \label{fig:contour}
\end{figure}

The two panels of Figure~\ref{fig:contour} display the $68\%$~CL~(upper pannel) and 95\%~CL~(lower panel) constraints in the $\kc \hspace{0.25mm}$--$\hspace{0.25mm} \kb$ plane for the four uncertainty scenarios introduced in Table~\ref{tab:errors}. Notice that the constraint arising in scenario~$S_1$~(blue contour) resembles the one shown Figure~3 of our letter, while the constraint corresponding to~$S_4$~(red contour) is slightly better than the LHC~Run~II region shown therein due to the smaller statistical uncertainty at the HL-LHC.

Profiling over~$\kappa_b$ we obtain the 95\%~CL limits on $\kappa_c$ reported in the last column of the Table~\ref{tab:errors}. One first observes that even under the assumption that the experimental and
theoretical systematic uncertainties will be the same as in the LHC Run II scenario, i.e.~scenario $S_4$, the resulting bounds on $\kc$ are not significantly worse than the limits corresponding to the scenario $S_1$. The relative similarity between the bounds on~$\kappa_c$ in the scenarios~$S_3$ and $S_4$, however, shows that an improved experimental systematic uncertainty can only be fully harnessed if theoretical errors are also reduced.

\begin{figure}[h!]
\centering
 \includegraphics[width=0.8\columnwidth]{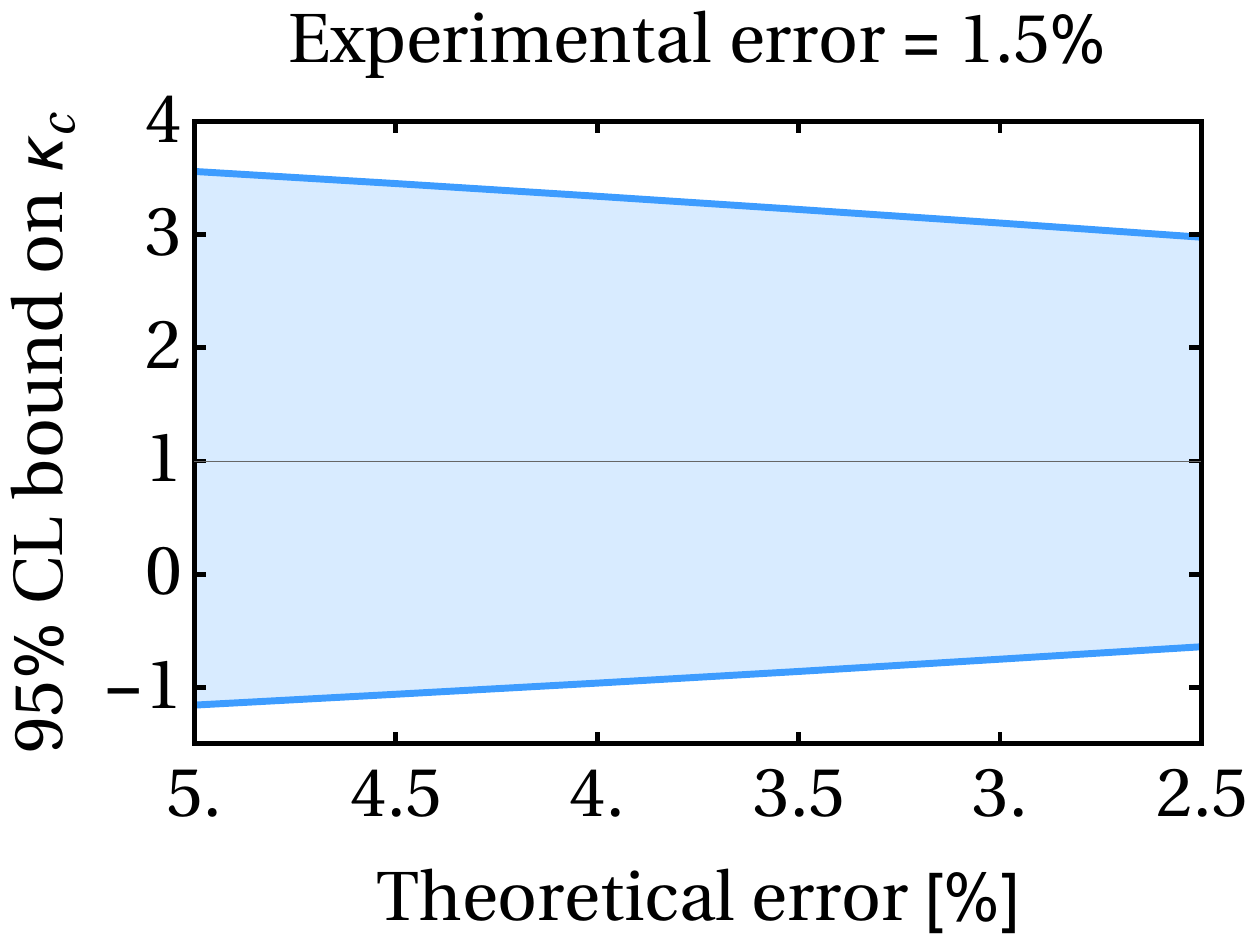}
\caption{95\%~CL bound on $\kappa_c$  as a function of the theoretical systematic error. The experimental systematic uncertainty  is fixed to $1.5\%$ in the plot, and the result has been profiled over $\kappa_b$. The shown band corresponds to $3 \, {\rm ab}^{-1}$ of integrated luminosity.}
  \label{fig:marginalized}
\end{figure}

The latter point is illustrated in Figure~\ref{fig:marginalized} which shows the upper and lower 95\%~CL limits on $\kappa_c$ as a function of theory error.  To obtain the plot we have profiled over~$\kappa_b$ and fixed the experimental uncertainty to $1.5\%$. One observes an approximately linear scaling of the bounds on~$\kappa_c$ with the variation of the theory error. This feature is generic as long as the experimental systematic uncertainty is smaller than the theoretical error.

\newpage 
\end{document}